\documentclass[intlimits,twoside,a4paper]{article}

\usepackage{amsmath,amssymb}
\usepackage{graphicx}

\usepackage[T2A]{fontenc}
\usepackage[cp1251]{inputenc}

\usepackage[eqsecnum]{cmpj3}

\usepackage{bm}
\usepackage{epsf}

\newcommand{\bea}{\begin{eqnarray}}
\newcommand{\eea}{\end{eqnarray}}

\issue{2018}{21}{1}{13001}
\doinumber{10.5488/CMP.21.13001}

\title[A simple ansatz for the study] {A simple ansatz for the study of velocity autocorrelation functions in fluids at different timescales}

\author[V.V.~Ignatyuk, I.M.~Mryglod, T.~Bryk]{V.V.~Ignatyuk, I.M.~Mryglod,  T.~Bryk}
 \address{Institute for Condensed Matter Physics of the National Academy of Sciences of Ukraine, \\
1 Svientsitskii St., 79011 Lviv, Ukraine}

\date{Received September 29, 2017}

\begin{document}

\maketitle

\begin{abstract}
A simple ansatz for the study of velocity autocorrelation
functions in fluids at different timescales is proposed. The
ansatz is based on an effective summation of the infinite
continued fraction at a reasonable assumption about convergence of
relaxation times of the higher order memory functions, which have a
purely kinetic origin. The VAFs obtained within our approach are
compared with the results of the Markovian approximation for
memory kernels. It is shown that although in the ``overdamped''
regime both approaches agree to a large extent at the initial and
intermediate times of the system evolution, our formalism yields
power law relaxation of the VAFs which is not observed at the
description with a finite number of the collective modes. Explicit
expressions for the transition times from kinetic to hydrodynamic
regimes are obtained from the analysis of the singularities of spectral functions 
in the complex frequency plane.


\end{abstract}

\section{Introduction\label{Introduct}}

The dynamics of many-body systems at various time scales remains as
one of the main problems of the non-equilibrium statistical
mechanics despite its long history and considerable achievements
obtained so far. Systematic studies of the time behaviour of
condensed matter systems started with the pioneering work by
Bogoluyubov \cite{Bogolyubov1}, where a concept of the
weakening correlations allowed one to reformulate a completely
unsolvable problem of the $N$-body dynamics into a much more
manageable task in terms of the correlation functions of lower
orders $s\ll N$.

Within such an approach, a time hierarchy of the phy\-si\-cal
stages, through which the system passed during its evolution
toward equilibrium (starting from the kinetic stage via
intermediate steps till the hydrodynamic one and beyond
\cite{MorozovBook}), appears quite naturally. The corresponding
time scales within that hierarchy can serve as a basis for
generalized hydrodynamics, in particular, in generalized
collective mode theory (GCM) \cite{GCM1,GCM2}, where the set of
dynamical variables for description of long- and short-time
correlations usually consists of the densities of conserved
quantities and their time derivatives up to a certain order. Using
a physically reasonable assumption in the Markovian approximation
\cite{GCM3} for the kinetic kernels of higher orders, one actually
comes to a simple mathematical problem, expressing collective
processes in the system in terms of the dynamic eigenmodes:
complex/real eigenvalues for propagating/relaxing processes and
corresponding eigenvectors which characterize contribution of
particular eigenmode to relevant correlations. A definite
advantage of the GCM is the fact that such a theory is a computer
adapted one: all the elements of the generalized dynamic matrix
could be expressed via the static correlation functions (SCF) and
corresponding relaxation times, which can be obtained by molecular
dynamic simulations \cite{GCM3,GCM4}.

Although the GCM can be extended by taking into account the
``ultraslow'' processes (defined by the time integrals of
corresponding densities \cite{Ome00,GCMT_slow,Bry10}), the
problems of account for slow structural relaxation
\cite{stuctur_relax} are usually approached by theories with non-local
coupling of dynamic variables. In the framework of the mode
coupling theory (MCT) \cite{MCT1}, a basic set of the dynamic
variables consists of higher products \cite{MCT3} rather than of
higher derivatives of the densities of conserved quantities. Like
in the GCM, a time/spatial dispersion of the kinetic kernels of
lower orders gives rise to some peculiarities of time correlation
functions (TCF). However, in contrast to the GCM, where the system
description at long times is a ``bottleneck'' of the theory, the
MCT approach mani\-fests its efficiency upon the hydrodynamic
stage of the system evolution yielding the $t^{-D/2}$ power law
for the TCFs \cite{long_tail1}. Recently \cite{MF_kernels2}, the
basic ideas of MCT have been used to obtain a mean field
approximation for the memory kernels, and to study a slow dynamics
of some soft matter systems like molecular liquids or colloid
suspensions. Such an approach was shown to yield reasonable
results in the whole time scale: from the ballistic motion of a
separate molecule to the diffusion regime.

Another approach that incorporates some advantages of both
theories mentioned above has been proposed in the papers
\cite{russians1, russians2}. The authors have used the continued
fraction method for the Laplace transforms of the VAFs and
reasonable approximations for the memory kernels at the stage of
the closure procedure with subsequent profound analysis of the VAF
peculiarities at the complex frequency plane. It was shown that
unlike the GCM case, characterized by a limited number of the
isolated poles, VAF shows the singularity manifold forming branch
cuts. The branch cuts were found to be separated from the real
axis by the well-defined ``gap''. The inverse value of the gap
width determines a duration of the transition period from the
short-time one-particle kinetics to the long-time collective
motion (hydrodynamics) with a typical power law relaxation $\sim
t^{-3/2}$ of the VAFs, which has been reported for the first time
and explained by Alder and Wainwright in \cite{Ald70}.
According to \cite{Ald70}, the long-time tails of VAF are
sustained by neighbour backflow created by redistribution of
momentum of the moving particle with neighbour particles.

Very recently, a microscopic interpretation of the long-time tails
of VAF was suggested from the analysis of its memory kernel
\cite{MHforMemKer}. It was shown, that the hydrodynamically added
mass, defined via memory kernel, has a negative sign, and the
backflow of neighbours tends to drag the particle in the direction
of its initial velocity, i.e., contributes negatively to the
friction.

There are also some other approaches to study the system dynamics,
which are not directly related to the concept of the weakening
correlations, or to the corresponding time hierarchy. Within the
Langevin formalism \cite{JCP}, a tagged particle is postulated to
interact with its neighbourhood as with a thermal bath, which
possesses certain spectral properties. In the framework of the
Mittag-Leffler generalization \cite{MittLefl2} of the Langevin
approach, a correlation function of the random noises behaves as a
stretched exponential for the short times and as an inverse power
law at the long-time scale. This allows one to describe different
dynamical regimes of the particle motion such as oscillations and
negative correlations of the VAFs, depending on the parameters of
the Mittag-Leffler function.

Summarizing all the above mentioned --- in spite of the considerable
achievements in exploration of dynamics of many-particle systems ---
there is still a lack of the unified self-consistent approaches
being able to describe the systems behaviour throughout all the
stages of evolution. An attempt to create such an approach is the
motivation of our paper. In our studies we i) present our
viewpoint on peculiarities of the fluid dynamics at different time
scales; ii) establish the reasons of the ``long tails'' origin in
the VAFs; and iii) investigate a crossover from the kinetics to
the hydrodynamics.

Our approach incorporates basic ideas of the GCM
\cite{GCM1,GCM2,GCM3}, MCT
\cite{MCT1,MCT3,long_tail1,MF_kernels2}, the formalism of
continued fractions for the time correlation functions
\cite{Bafile1,Bafile2}, and the method of the TCFs analysis on the
plane of complex frequency elaborated in
\cite{russians1,russians2}. To be specific, we consider the VAF as
a particular case of the more general TCF and use a continued
fraction representation for its Laplace transforms. In such an
approach, there appears a set of SCFs dealt with a force acting on
the particle and its higher derivatives. These SCFs serve as the
control parameters $\Gamma_i$ to be evaluated, for instance, from
computer simulations. A cornerstone of our scheme is an effective
summation of infinite continued fractions, based on a quite
reasonable (and physically grounded) assumption about a
convergence of relaxation times of the higher modes
\cite{our_assump1,our_assump2} of purely kinetic origin. This
allows us to obtain an explicit expression for the memory kernel
of the highest order, which turned out to be oscillatory in time
representation and decays as $t^{-3/2}$.

We show that at a wide enough domain of the control parameters
$\Gamma_i$, the VAFs, calculated within our approach, almost
coincide at the initial and intermediate times with those
obtained in the framework of the finite modes approximation.
However, in contrast to the GCM, which implies a description by
the set of exponents and does not permit the ``long tails'' to
appear at the hydrodynamic stage of the system evolution, we
obtain a specific $\sim t^{-3/2}$ relaxation (or even slower decay
$\sim t^{-1/2}$ at a certain critical value of the parameter
$\Gamma_i$). At the same time, if we fix a level $s$, where the
summed up kinetic kernel is located, we can provide the frequency
moments up to the $2s$-th order as well as the zeroth time moment
of the VAF to obey the ``sum rules'' \cite{GCM2,GCM3,GCM4}, like
it holds within the GCM framework.

In addition, we show that a transition time from the exponential
relaxation at the intermediate stage of the system evolution to
the power law decay at the hydrodynamic stage depends on the
distance of the corresponding singularities of the spectral
functions from the coordinate origin. Our results are in a close
agreement with the conclusions of
\cite{russians1,russians2}, though the starting point of our
approach differs from the basic assumptions of the cited papers.

  A structure of the paper is as follows. In section~\ref{secII}, we
 briefly recall the method of continued fractions for the VAF presentation,
  and calculate the memory function of the highest order at the
  assumption mentioned above. In section~\ref{secIII}, we evaluate the
  spectral functions both by the
  effective summation of the continued fraction and in the finite
  modes approximation.
    In section~\ref{secIV}, an analysis of the VAFs dynamics is performed.
The VAFs are shown to behave in a solid-like, gas-like or
liquid-like manner depending on the control parameters $\Gamma_i$.
A special attention will be paid to the
  case when the fluid behaves as an effective $1\text{D}$ system with
  typical $\sim t^{-1/2}$ decay at long times.
  In
  section~\ref{secV}, we investigate in detail a transition of the VAFs to the
  power law asymptotics.  Finally, we draw the conclusions in
section~\ref{secVI}.

\section{Continued fraction representation for velocity autocorrelation function\label{secII}}

Let us consider the normalized VAF \bea\label{VAF}F_v(t)=\langle
\textbf v_i\exp(-\textrm i L_N t)\textbf v_i\rangle/\langle \textbf
v_i\textbf v_i\rangle,\eea where $\textrm i L_N$ means the Liouville
operator. For the system of structureless particles interacting
via the central force potential it has a very simple form
\cite{MorozovBook},
being dependent on positions and momenta of all the particles.

The Laplace-transformed VAF as a formal solution of the
generalized Langevin equation reads
$$
\tilde F_v(z)=\frac{1}{z+\tilde{\phi}_1(z)}~,
$$
where $\phi_0(z)$ is the Laplace-component of the lowest-order
memory kernel which should contain all the information on the
dissipation processes in a fluid, an explicit expression for
which should be obtained from the higher-order memory kernel
$\phi_s(z)$, which satisfies a recurrence relation \cite{Boo}
\bea\label{recurrV}
\tilde{\phi}_{s-1}(z)=\frac{\Gamma_{s-1}}{z+\tilde{\phi}_{s}(z)}\,,\eea
where $\Gamma_{s-1}$ is the relevant SCF. It is possible to
present the equation for the Laplace transform $\tilde F_v(z)$ of
VAF (\ref{VAF}) as an infinite continued fraction
\bea\label{VAFfrac} \tilde F_v(z)=
\displaystyle\displaystyle\frac{1}{z+\displaystyle\frac{\Gamma_1}{z+\displaystyle\frac{\Gamma_2}{z+\displaystyle\frac{\Gamma_3}{z+\cdots}}}}\,.
\eea All the parameters $\Gamma_i$ entering (\ref{VAFfrac})
are expressed according to a general definition \cite{Boo} via the
``force-force'' SCFs and its higher derivatives. The lowest order
SCFs can be presented as follows: \bea\label{Gamma12}
\Gamma_1=\frac{\langle\mathbf F_i \mathbf
F_i\rangle}{\langle\mathbf v_i \mathbf v_i\rangle}=
\frac{\beta}{3m}\langle\mathbf F_i \mathbf
F_i\rangle,\qquad \Gamma_2=\frac{\langle{\textrm i L_N\mathbf
F_i \textrm i L_N\mathbf F_i}\rangle}{\langle\mathbf F_i \mathbf
F_i\rangle}-\Gamma_1.
 \eea

To evaluate the static correlators $\Gamma_1$, one should obtain a
pair correlation function of the fluid at a given thermodynamic
point. The SCF $\Gamma_2$ would require the knowledge of even higher
order correlation functions, which yields the problem completely
unmanageable. Therefore, such a structural analysis of the fluid
could not be considered as a promising one, and reliable
results for VAFs as well as for higher order SCFs $\Gamma_i$ should
be obtained only by MD simulation. Any theoretical
approximation should be compared with the MD data to verify an
efficiency of the chosen approach.

The infinite continued fraction (\ref{VAFfrac}) can be rewritten
\cite{Bafile1} as a weighted sum \bea\label{sumZ} \tilde
F_v(z)=\sum\limits_{j=1}^{\infty} \frac{I_j}{z-z_j}\,, \eea where
$I_j$ denote the amplitudes of particular $j$-th mode of the
fluid, whereas $z_j$ mean the mode frequencies. In the framework
of GCM \cite{GCM3,mryglod98}, these values can be calculated by
solving the eighenvalues/eighenvectors problem for the generalized
dynamic matrix of the infinite dimension, while in the other
approach \cite{Bafile1,Bafile2} they are taken as fitting
parameters. In fact, an evaluation of all the higher order
correlators $\Gamma_i$ is not a realistic task. Therefore, for
practical purposes one should truncate the series (\ref{sumZ}),
retaining only a limited number $M$ of terms. In accordance
with the basic concept of GCM, it corresponds to the $M$ modes
contribution to the VAF.

For instance, limiting ourselves by two terms in (\ref{sumZ})
(this corresponds to the Markovian approximation
$\tilde{\phi}_2(z)\approx \tilde{\phi}_2(0)$ for the second order
memory kernel), and writing down \bea\label{phiMarkov}
\tilde{\phi_1}(z)\approx\frac{\Gamma_1}{z+\tilde{\phi}_2(0)}\,, \eea
we can reproduce equation~(15) of \cite{GCM4} for the VAF of a
simple fluid. Such a model admits the existence of either two
purely relaxing or two propagating modes in the VAF dynamics,
depending on the relation between two typical times of the system
evolution. These times are: i) duration of the hydrodynamic
processes connected with the self-diffusion, and ii) the inverse
Einstein frequency.

We have already mentioned that construction of the exact
representation of VAF as an infinite continued fraction
(\ref{VAFfrac}) seems to be an unrealistic problem. We note that a
truncation of the series (\ref{sumZ}) even at a large number of terms
may perturb the VAF's dynamics. This distortion can be
insignificant (allowing one to describe the system dynamics at small
and intermediate times in the GCM framework) or essential (not
allowing one to study a hydrodynamic regime of the system evolution).
However, even in the first case, the issue of truncation of the series
(\ref{sumZ}) to approximate the VAF by a finite number of modes
as accurately as possible should be solved separately for each particular system 
and for each particular thermodynamic point \cite{Bafile2}. As regards the second case,
from a strictly mathematical viewpoint there is not another reason
of the power law behaviour of the VAFs at long times than a
contribution of \textit{the infinite number} of terms in
(\ref{VAFfrac}). In the MCT approach \cite{MCT1,MCT3}, which is
not directly related to a truncation of the continued fractions,
the system dynamics is modelled at the long times but could fail
at the other timescales.

Therefore, we propose to adopt another approach for an approximate
evaluation of the infinite continued fraction (\ref{VAFfrac}). The
Markovian approximation of the lowest order kinetic kernel is
known to determine \cite{mryglod98} the relaxation time $\tau_1$
of the corresponding hydrodynamic excitation by a simple relation
$\tilde{\phi}_1(0)=\tau_1^{-1}$. Introducing in a similar manner
the $s$-th order relaxation times, and using a fact
\cite{our_assump2,mryglod98} that the relaxation times of the
higher modes become comparable with each other and tend to a
certain fixed value, which is of a purely kinetic origin, we can put
this assumption for large enough $s$ into an explicit form,
\bea\label{phiS_equal}
\tilde{\phi}_s(0)\approx\tilde{\phi}_{s+1}(0)\approx\tilde{\phi}_{s+2}(0)=\ldots .
\eea Keeping in mind equation~(\ref{phiS_equal}) and making use of the
relation (\ref{recurrV}), where the condition
$\tilde{\phi}_s(0)=\tilde{\phi}_{s+1}(0)$ is used, one can
obtain a simple expression for the $s$-th order memory function in
the Markovian approximation,
\bea\label{phiS_GammaS}\tilde{\phi}_s(0)\approx\sqrt{\Gamma_s}\,.\eea
It also follows from equations~(\ref{recurrV})--(\ref{phiS_equal})
that all the higher order SCFs $\Gamma_i$, $i=\{s,s+1,\ldots\}$,
are almost equal to each other, \bea\label{GammaS}
\Gamma_s=\Gamma_{s+1}=\Gamma_{s+2}=\ldots, \eea whereas the
remaining SCFs with $2i+1<s$ yield the recurrence relations for
the corresponding relaxation times, \bea\label{recurrTAU}
\frac{1}{\tau_{2 i}}=\frac{1}{\tau_{2
i-1}}\sqrt{\frac{\Gamma_{2i}}{\Gamma_{2i+1}}}\,,\qquad
\frac{1}{\tau_{2 i+1}}=\frac{1}{\tau_{2
i}}\sqrt{\frac{\Gamma_{2i+2}}{\Gamma_{2i+1}}}\,.\eea One can see
from equation~(\ref{recurrTAU}) that there are, actually, two sets of
relations for the odd and the even $i$, separately. It is known
from the theory of continued fractions \cite{Campos} as well as
from the analysis of fluid dynamics \cite{our_assump2} that
$\tau_i$ oscillates around a certain value $\tau^*$, approaching
this asymptotics from below (the odd order relaxation times) or
above (the even order relaxation times). Moreover, the difference
between $\tau_{2i}$ and $\tau_{2i+1}$ decreases rapidly with
increasing $i$.

If one assumes $\tau_s$ to reach its asymptotics instantly
[this, in fact, corresponds to approximation~(\ref{phiS_equal})],
the recurrence relation (\ref{recurrV}) reduces to a bilinear form
\bea\label{recurrPHI}
\tilde{\phi}_s(z)=\frac{\Gamma_s}{z+\tilde{\phi}_s(z)}\,, \eea hence,
an explicit expression for the $s$-th order memory kernel can
be easily obtained, \bea\label{phiS_result}
\tilde{\phi}_s(z)=-\frac{z}{2}+\sqrt{\frac{z^2}{4}+\Gamma_s}\,. \eea
This result looks quite interesting if one performs the inverse
Laplace transformation to (\ref{phiS_result}), \bea\label{phiS_t}
\phi_s(t)\equiv\frac{1}{2\piup \textrm i}\lim\limits_{\epsilon\to
0}\!\!\int\limits_{\epsilon-\textrm i\infty}^{\epsilon+\textrm i\infty}\!\!\! \textrm e^{z
t}\tilde{\phi}_s(z)\rd z
=\sqrt{\Gamma_s}\,\,\frac{J_1\big(2\sqrt{\Gamma_s}t\big)}{t}\,. \eea It
means that the $s$-th order me\-mo\-ry kernels decay
non-monotonously at long times as $t^{-3/2}$ since the Bessel
function of the first order $J_1$, entering equation~(\ref{phiS_t}),
obeys the well known asymptote $ J_1(x\gg 
1)=\sqrt{\frac{2}{\piup x}}\cos\left(x-3\piup/4\right)$.

Equations~(\ref{phiS_result})--(\ref{phiS_t}), obtained in the framework
of the rigorous approach using just one physically reasonable
assumption (\ref{phiS_equal}), are the cornerstones for our
further study of the VAFs dynamics. In the next section we
consider the spectral functions with various values $s$, at which
the condition (\ref{phiS_equal}) starts to hold true. These
results are compared with the finite continued fraction approach,
for which the corresponding Markovian approximation yields the
expression \bea\label{phiS_Mark} \tilde{\phi}_{s-1}(z)\approx
\frac{\Gamma_{s-1}}{z+\tilde{\phi}_s(0)}
=\frac{\Gamma_{s-1}}{z+\sqrt{\Gamma_s}} \eea for the $(s-1)$-th
order memory function.

Later on, we perform an inverse Fourier transformation to obtain a
time representation $F_v(t)$ of the corresponding VAFs and to show
that even two levels of the hierarchy, at which the summed up
kinetic kernel (\ref{phiS_result}) is located, allow the VAFs
to mimic a ``gas-like'', ``liquid-like'' or ``solid-like''
behaviour of the simple fluid depending on the interplay between
the SCFs $\Gamma_i$. An essential feature is the appearance of
long time tails of the VAFs, which cannot be obtained within GCM.

\section{Spectral functions\label{secIII}}

Before representing the VAF by means of the summed up continued
fraction $(\ref{phiS_result})$, one can ask a natural question:
what level $s$ can be considered as a sufficient one for the
higher order memory kernels $\tilde{\phi}_s(0)$ in order to reach their
asymptotic value and the relation (\ref{phiS_equal}) to be true?
This would require sophisticated calculations of the sequence of
SCFs by computer simulations. However, to obtain some analytical
estimates, in this section we assume that the condition
(\ref{phiS_equal}) is valid starting from quite low values $s\geqslant
2$.

Let us consider the spectral function (SF) $\tilde
F_v(\omega)=\Re\big[\tilde F_v(z=\textrm i\omega)\big]$, defined as a real
part of the correspond\-ing continued fraction (\ref{VAFfrac})
taken at the imaginary frequency. Hereafter, we replace the label~$v$ in the expressions for SFs (and for VAFs) by the subscript
$\alpha$ which denotes the chosen level of approximation (for details see
below).

If one supposes the following relation
\bea\label{G2}\Gamma_2=\Gamma_3=\ldots=\Gamma_s=\ldots \eea to be
valid, it is straightforward to evaluate SF $\tilde F_2(\omega)$:
\bea\label{spectr2} \tilde
F_2(\omega)=\frac{\Gamma_1\sqrt{\Gamma_2-\omega^2/4}}{(\Gamma_2-\Gamma_1)\omega^2+\Gamma^2_1}\,.
\eea It is also instructive to compare the result (\ref{spectr2})
with the expression 
\bea\label{spectrM2} \tilde
F_{\text{M2}}(\omega)=\frac{\Gamma_1\sqrt{\Gamma_2}}{\omega^4+(\Gamma_2-2\Gamma_1)\omega^2+\Gamma_1^2}
\eea 
for the SF, obtained by a truncation of the continued
fraction at the level $s=2$ [or, which is the same, after the
Markovian approximation
$\tilde{\phi}_2(z)\approx\tilde{\phi}_2(0)$ for the memory kernel
of the 2-nd order]. Hereafter, we use an extra subscript ``M'' to
denote the above mentioned Markovian approximation.

It is useful to compare the expressions
(\ref{spectr2})--(\ref{spectrM2}) for the corresponding SFs. First
of all, the former spectral function vanishes at the cut-off
frequency $\omega_{\textrm{c}}=2\sqrt{\Gamma_2}$, while the latter tends to
zero asymptotically as $1/\omega^4$. Secondly, both SFs are even
functions of frequency, and behave as $\lim_{\omega\to 0}\rd\tilde
F_{\alpha}(\omega)/\rd\omega=0$, here $\alpha$ denotes 2 or M2.
Thirdly, in the domain of frequencies $\omega\ll\omega_{\textrm{c}}$ and at
$\Gamma_1\ll\Gamma_2$, the relation $\tilde
F_2(\omega)\approx\tilde F_{\text{M2}}(\omega)$ holds true [to put it
more precisely, an exact relation $\tilde F_2(0)=\tilde F_{\text{M2}}(0)$
is valid, see discussion in section~\ref{secIV}].

At last, in the vicinity $\varepsilon$ of the
cut-off frequency $\omega_{\textrm{c}}$, the SF (\ref{spectr2}) behaves as
\bea\label{sqrtEps} \tilde
F_2(\omega_{\textrm{c}}-\varepsilon)\approx\frac{\Gamma_1\Gamma_2^{1/4}}{(\Gamma_1-2\Gamma_2)^2}\sqrt{\varepsilon}.
\eea

The square root dependence (\ref{sqrtEps}) resembles the
well-known result from the MCT \cite{MCT3,long_tail1}. However, in
the MCT approach, a non-analytical dependence $\sqrt\omega$ of the
SFs, generating the long time tails $\sim t^{-3/2}$ of the VAFs,
occurs at the low frequency domain rather than at $\omega_{\textrm{c}}$.

A similar structure of the SF can be obtained at the assumption
\bea\label{G3}\Gamma_3=\Gamma_4=\ldots=\Gamma_s=\ldots, \eea when
all the relaxation times $\tau_s$, starting from $s=3$, are
supposed to be the same (compare with the condition (\ref{G2})).
In this case, the spectral function $\tilde F_3(\omega)$, obtained
by summing up the continued fraction for the 3-rd order memory
kernel, can be presented as follows: \bea\label{spectr3} \tilde
F_3(\omega)=\frac{\Gamma_1\Gamma_2\sqrt{\Gamma_3-\omega^2/4}}{(\Gamma_3-\Gamma_2)\omega^4+
\big[\Gamma_2(\Gamma_1+\Gamma_2)-2\Gamma_1\Gamma_3
\big]\omega^2+\Gamma_1^2\Gamma_3}\,, \eea while its counterpart in
the modes approximation looks as
\bea\label{spectrM3} \tilde F_{\text{M3}}(\omega)=
\frac{\Gamma_1\Gamma_2\sqrt{\Gamma_3}} {\omega^6-\big[
2(\Gamma_1+\Gamma_2)-\Gamma_3\big]\omega^4+\big[
(\Gamma_1+\Gamma_2)^2-2\Gamma_1\Gamma_3
\big]\omega^2+\Gamma_1^2\Gamma_3}\,. \eea

Like in the the cases (\ref{spectr2})--(\ref{spectrM2}), the two
SFs are close to each other, $\tilde F_3(\omega)\approx\tilde
F_{\text{M3}}(\omega)$, in the domain of low frequencies
$\omega\ll\omega_{\textrm{c}}$ and at
$\Gamma_3\gg\mbox{max}\{\Gamma_1,\Gamma_2\}$.

It should be stressed that there is a restriction from below for
the minimal value $\Gamma_s^{(\textrm{min})}$ of the highest order SCF (in
our case, $s=2$ or 3), which follows from the sum rule
\bea\label{sumRule0}
\frac{1}{2\piup}\int\limits_{-\infty}^{\infty}\tilde
F_s(\omega)\rd\omega\equiv\frac{1}{2\piup}\int\limits_{-\omega_{\textrm{c}}}^{\omega_{\textrm{c}}}\tilde
F_s(\omega)\rd\omega=1. \eea Equation~(\ref{sumRule0}), in fact, denotes
the initial value of the VAF, which is equal (by definition) to
the unity. A direct integration of (\ref{spectr2}) shows that at
a fixed value $\Gamma_1$, the sum rule (\ref{sumRule0}) holds
true for all $\Gamma_2\geqslant\Gamma_2^{(\textrm{min})}$, where
$\Gamma_2^{(\textrm{min})}=\Gamma_1/2$.

In a similar way, the value
$\Gamma_3^{(\textrm{min})}=(\Gamma_1+2\Gamma_2)/4$ can be defined for a
fixed lower order SCFs $\Gamma_1$, $\Gamma_2$ in
equation~(\ref{spectr3}), and the sum rule (\ref{sumRule0}) is
satisfied for $\Gamma_3\geqslant\Gamma_3^{(\textrm{min})}$. In this context, our
results differ from those [see equations~(\ref{spectrM2}),
(\ref{spectrM3})] obtained within the modes approximation: in the
latter case, no relation between various $\Gamma_i$ is needed for
the sum rule (\ref{sumRule0}) to be satisfied.

We should emphasize that so far the minimal values
$\Gamma_s^{(\textrm{min})}$ appear in a strictly mathematical way due to
the requirement of the sum rule (\ref{sumRule0}). Later on, we
will try to attribute a more physical meaning for
$\Gamma_s^{(\textrm{min})}$ (as well as for $\omega_{\textrm{c}}$), when studying the
time behaviour of the corresponding VAFs and discussing the
process of the long tails formation in section~\ref{secV}.

At the end of this section, we would like to present an
expression for spectral weight functions~$J(\omega)$, which
determine the generalized friction coefficient
\bea\label{Gfriction}
\gamma(t)=\frac{1}{\piup}\int\limits_0^{\infty}
\frac{J(\omega)}{\omega}\cos\omega t\,\rd\omega ,\eea entering the
Langevin equation for the tagged particle of the fluid, which is
supposed to interact with its neighbourhood as a collection of the
$N$ bath modes \cite{JCP}. Identifying \cite{MittLefl2} the
generalized friction coefficient with the 2-nd order memory kernel
(\ref{phiS_t}) and taking into account equation~(\ref{Gfriction}), one
can obtain an expression for the spectral weight function as
follows:
 \bea\label{OhmC}
J(\omega)=\frac{4\Gamma_2}{\piup\omega_{\textrm{c}}^2}\omega\sqrt{\omega_{\textrm{c}}^2-\omega^2}\,.
\eea

It is seen from equation~(\ref{OhmC}) that the fluid corresponds to the
Ohmic system \cite{Leggett}, since in the low frequency domain the
spectral weight function behaves as $J(\omega)\sim\omega$. The
obtained result looks quite expected. However, there is a cut-off
frequency $\omega_{\textrm{c}}$ at which $J(\omega)$ abruptly vanishes. In
this context, equation~(\ref{OhmC}) differs from the commonly accepted
form for the spectral weight functions, which are usually
expressed as \bea\label{Jomega1}
J(\omega)\sim\omega^n\exp(-\omega/\omega_{\textrm{c}})\eea with
$0\!<\!n\!<\!1$ for the sub-Ohmic systems, $n=1$ for the Ohmic
ones, and $n>1$ for the super-Ohmic coupling \cite{Leggett}.

It is widely believed that the low frequency asymptote determines
the long time behaviour of the corresponding TCF, and the cut-off
frequency in the exponent of (\ref{Jomega1}) can just slightly
change the shape of TCF. However, in the theory of solids, it is
reasonable \cite{JCP1992,myJCP} to consider the upper cut-off
frequency $\omega_{\textrm{c}}$, associated, for instance, with the Debye
frequency $\omega_{\textrm{D}}$, and to put $J(\omega)=0$ at
$\omega>\omega_{\textrm{c}}$, like it happens in the case under
consideration. An advantage of our approach consists in the fact
that the expression~(\ref{OhmC}) has been obtained rigorously
using the only assumption~(\ref{G2}) for the SCFs of higher
orders, and the cut-off frequency $\omega_{\textrm{c}}$ has appeared in a
natural way (not as a fitting parameter). Although from the
mathematical standpoint any truncation of the spectral weight
function at the finite frequency $\omega_{\textrm{c}}$ generates an oscillating
power law behaviour of the corresponding TCF at long times, it is
in our approach that a correct asymptotics $t^{-3/2}$ of VAF is ensured.

\section{Time behaviour of the VAFs\label{secIV}}

Expressions (\ref{spectr2})--(\ref{spectrM2}) and
(\ref{spectr3})--(\ref{spectrM3}) allow us to study the frequency
dependence of SFs in a broad domain of parameters $\Gamma_s$ as
it has been discussed in section~\ref{secIII}. However, it would be
more instructive to perform the inverse Fourier transformation of
the above mentioned expressions and to consider the dynamics of
the corresponding VAFs.

One important remark is to the point. In this section, we consider
various values of $\Gamma_s$ regardless of their relations to the
fluid at a particular thermodynamic point. We reckon our task in
presentation of different regimes of the fluids dynamics, modelled
in the framework of our approach. Certainly, all these SCFs need a
verification by computer simulations. Moreover, it may turn out
that some combinations of $\Gamma_s$ can even be non-representative
for real fluids. Nevertheless, preliminary estimations show
that the basic features of the fluids dynamics such as the short
time behaviour, the transition regimes, formation of the long
tails, etc., presented in this paper, are also valid with
$\Gamma_s$, obtained in a strict way (by computer simulations at
various temperatures and densities).

\begin{figure}[b!]
\centerline{\includegraphics[height=0.21\textheight,angle=0]{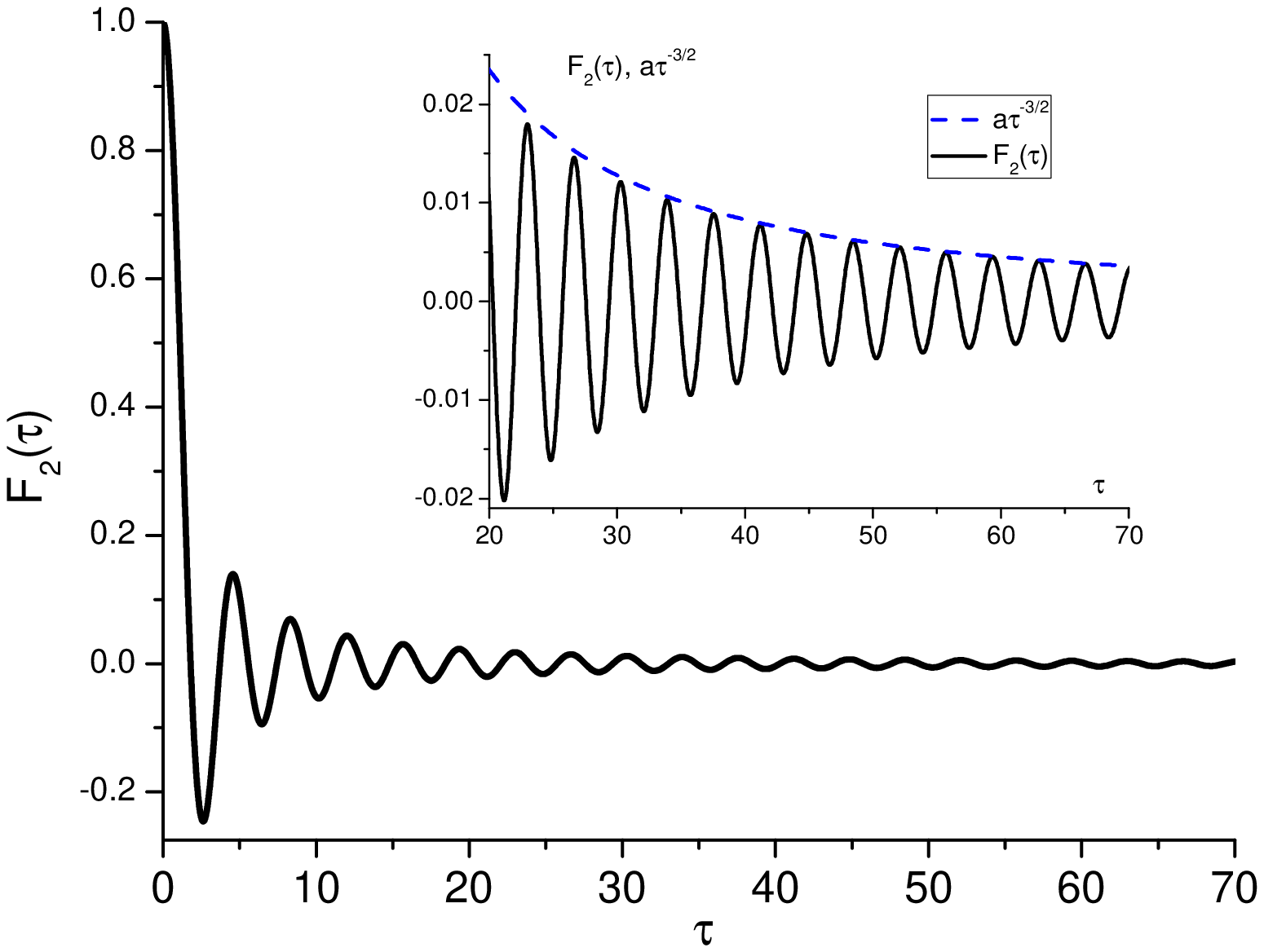}
 \includegraphics[height=0.21\textheight,angle=0]{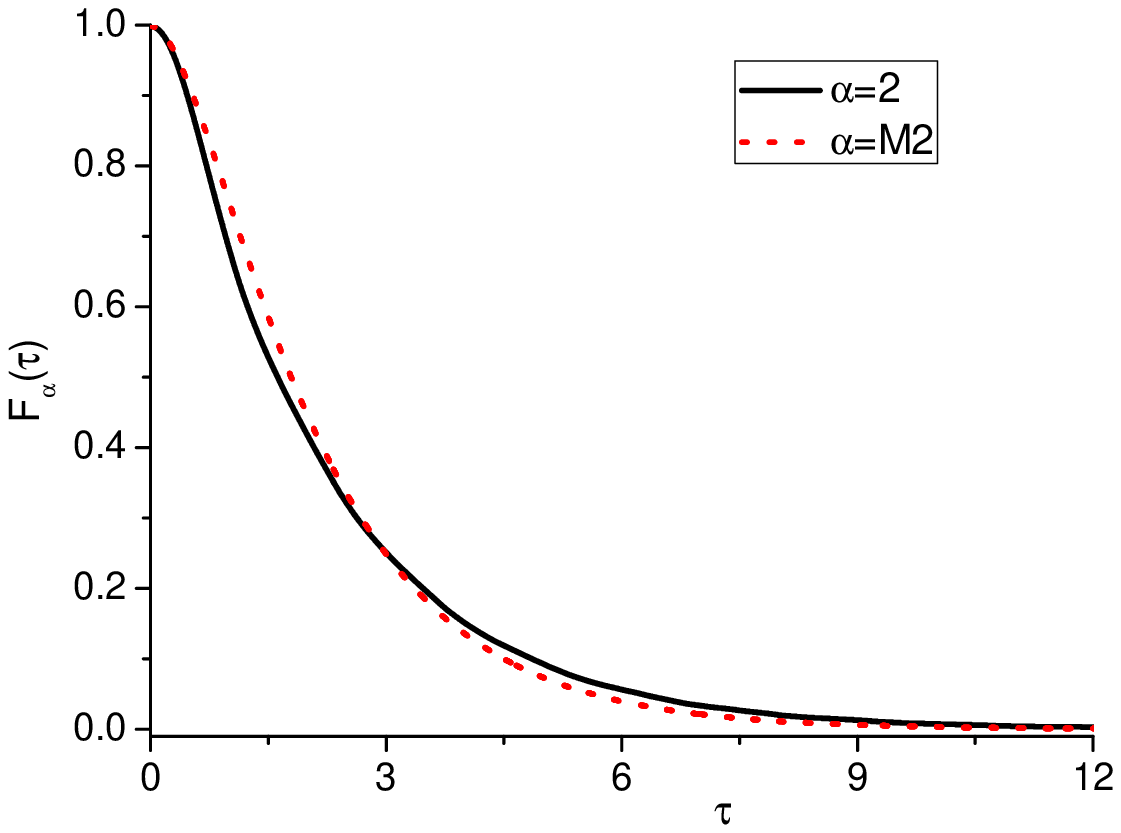}}
 \caption{(Colour online) Time evolution of the VAF $F_2(\tau)$
(solid line) at $\Gamma_1\tau^2_{\textrm{LJ}}=1$,
$\Gamma_2\tau^2_{\textrm{LJ}}=0.75$ (left-hand panel) and
$\Gamma_1\tau^2_{\textrm{LJ}}=1$, $\Gamma_2\tau^2_{\textrm{LJ}}=5$ (right-hand panel).
The dotted line corresponds to the VAF in the Markovian
approximation (labelled by $\alpha=\text{M2}$). In the inset: the VAF at
long times (solid line) and its asymptotics $a\tau^{-3/2}$ (dashed
line) with $a=2.1$. The dimensionless value $\tau=t/\tau_{\textrm{LJ}}$ is
defined via the typical time
$\tau_{\textrm{LJ}}=\sigma_{\textrm{LJ}}\sqrt{m/\varepsilon_{\text{LJ}}}$ of the
Lennard-Jones system.}
\label{fig1}
\end{figure}

We begin our investigation using approximation (\ref{G2}). In
figure~\ref{fig1} we present the time be\-ha\-vior of the VAFs calculated at
two different parameters $\Gamma_2$ whereas $\Gamma_1$ is chosen
to be fixed. It could be said (see the left-hand panel of figure~\ref{fig1}) that
at small values of $\Gamma_2$, the VAF shows a solid-like
\cite{VAFbook} behaviour since there are well defined oscillations
in the whole time domain, which decay as $a t^{-3/2}$ at long
times, as it is shown by the dashed line. The parameter $a$ is
nothing but a fraction in the expansion (\ref{sqrtEps}) divided by
$\sqrt{\piup}$. Obviously, we are not dealing with a real solid: the
oscillations are damped, and the area under the curve, which
determines the self-diffusion coefficient, differs from zero [it
also follows from the inequality $\tilde F_2(0)\ne 0$, see
equation~(\ref{spectr2})].

At a larger value of $\Gamma_2$, the VAF behaves in a gas-like
manner. In a gas, the molecules are not confined by intermolecular
bonding, and the VAF decays without any oscillation. In our case,
there is a smooth relaxation of $F_2(t)$ at the initial and at
intermediate times. It is worthy to point out that the two modes
approximation $F_{\text{M2}}(t)$ yields a very similar result (see the
dotted line in figure~\ref{fig1}), as it has been mentioned in
section~\ref{secIII}.

This close resemblance becomes even more reasonable if one notices
that both approaches ensure an equality of the frequency moments
$\langle\omega_{\alpha}^m\rangle=\frac{1}{2\piup}\int_{-\infty}^{\infty}\omega^m
\tilde F_{\alpha}(\omega) \rd\omega$ of the low orders,
\bea\label{some_M}\nonumber
\nonumber & \left\langle\omega^0_2\right\rangle=\left\langle\omega^0_{\text{M2}}\right\rangle=1,\\
&\left\langle\omega^2_2\right\rangle=\left\langle\omega^2_{\text{M2}}\right\rangle=\Gamma_1\,,\nonumber\\
&\left\langle\omega^4_2\right\rangle=\Gamma_1(\Gamma_1+\Gamma_2),\qquad
\left\langle\omega^4_{\text{M2}}\right\rangle=\infty. \eea It should be
also mentioned that the higher order frequency moments obtained
within the Markovian approximation diverge [see the last row in
equation~(\ref{some_M})] whereas all the frequency moments in our
scheme are finite being calculated by integration over the finite
interval $[-\omega_{\textrm{c}},\omega_{\textrm{c}}]$.

A similar equality, \bea\label{tau_M}
\left\langle\tau^0_2\right\rangle=\left\langle\tau^0_{\text{M2}}\right\rangle=\frac{\sqrt{\Gamma_2}}{\Gamma_1}\,,\eea 
can be obtained for the zeroth time moment
$\left\langle\tau_{\alpha}^0\right\rangle= \int_{0}^{\infty}
F_{\alpha}(t) \rd t$, which defines a self-diffusion coefficient of
the fluid. The higher order time moments (starting from $s\geqslant 2$),
obtained in our approach, diverge due to the long tail formation
(see the inset in the left-hand panel of figure~\ref{fig1}).

If $\Gamma_2$ is much greater than  $\Gamma_1$, as it occurs in
the diluted fluids and is confirmed by MD, the power law regime
can be formally set at the infinitely large times, which are not
accessible by the computer simulation [for estimation of the
transition times see equation~(\ref{tauS}) in section~\ref{secV}]. The
oscillations with a vanishing amplitude along with the envelope
curve $a t^{-3/2}$ (not shown in the right-hand panel of figure~\ref{fig1})
cannot, in fact, be observed. Therefore, there is not a
controversy between the predicted power law behaviour of the VAFs,
modulated by the high frequency $\omega_{\textrm{c}}$ at long times, and the
non-reversed motion of the gas particles at the whole time domain.

\begin{figure}[b!]
\centerline{\includegraphics[height=0.20\textheight,angle=0]{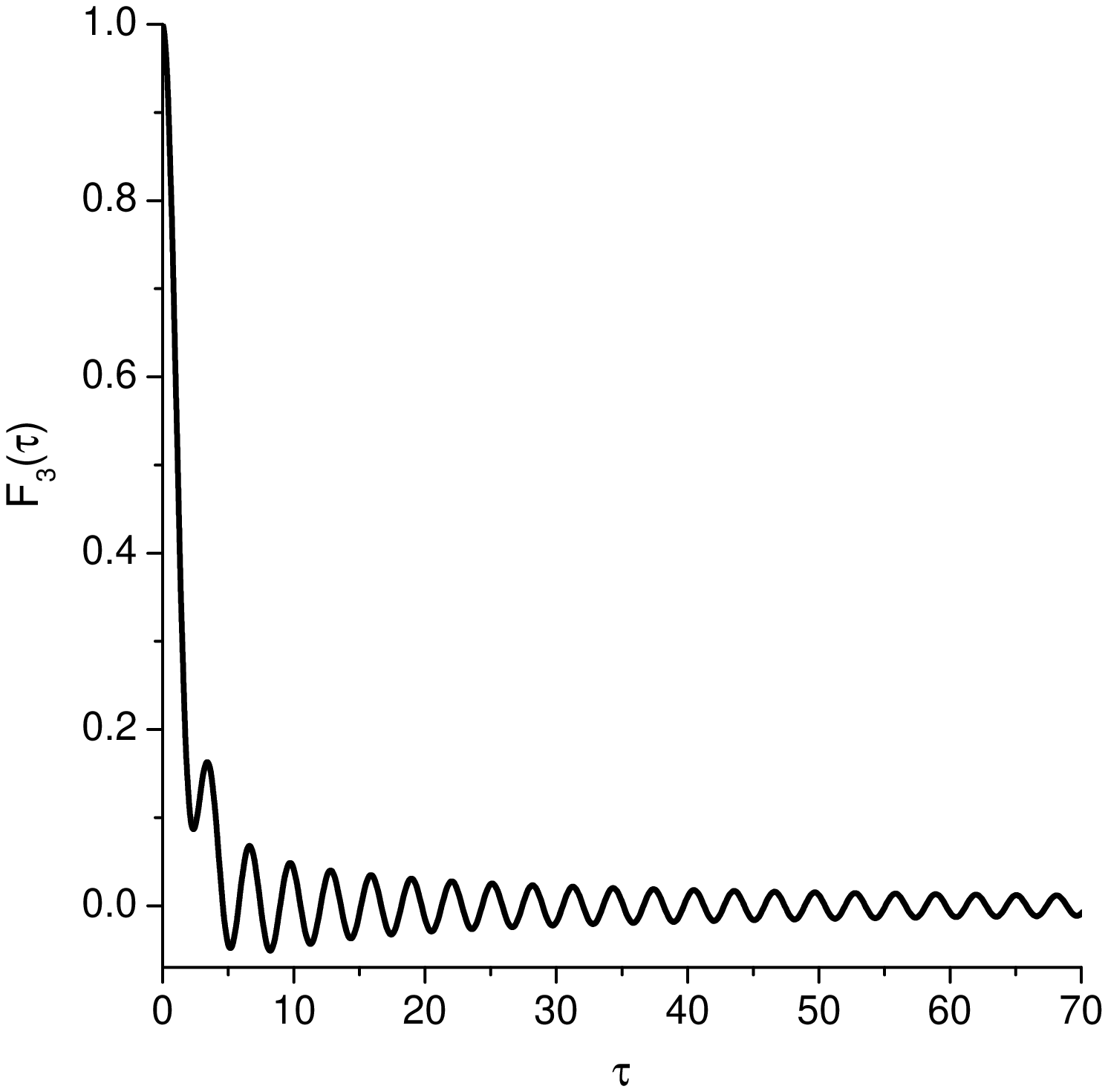}
\includegraphics[height=0.20\textheight,angle=0]{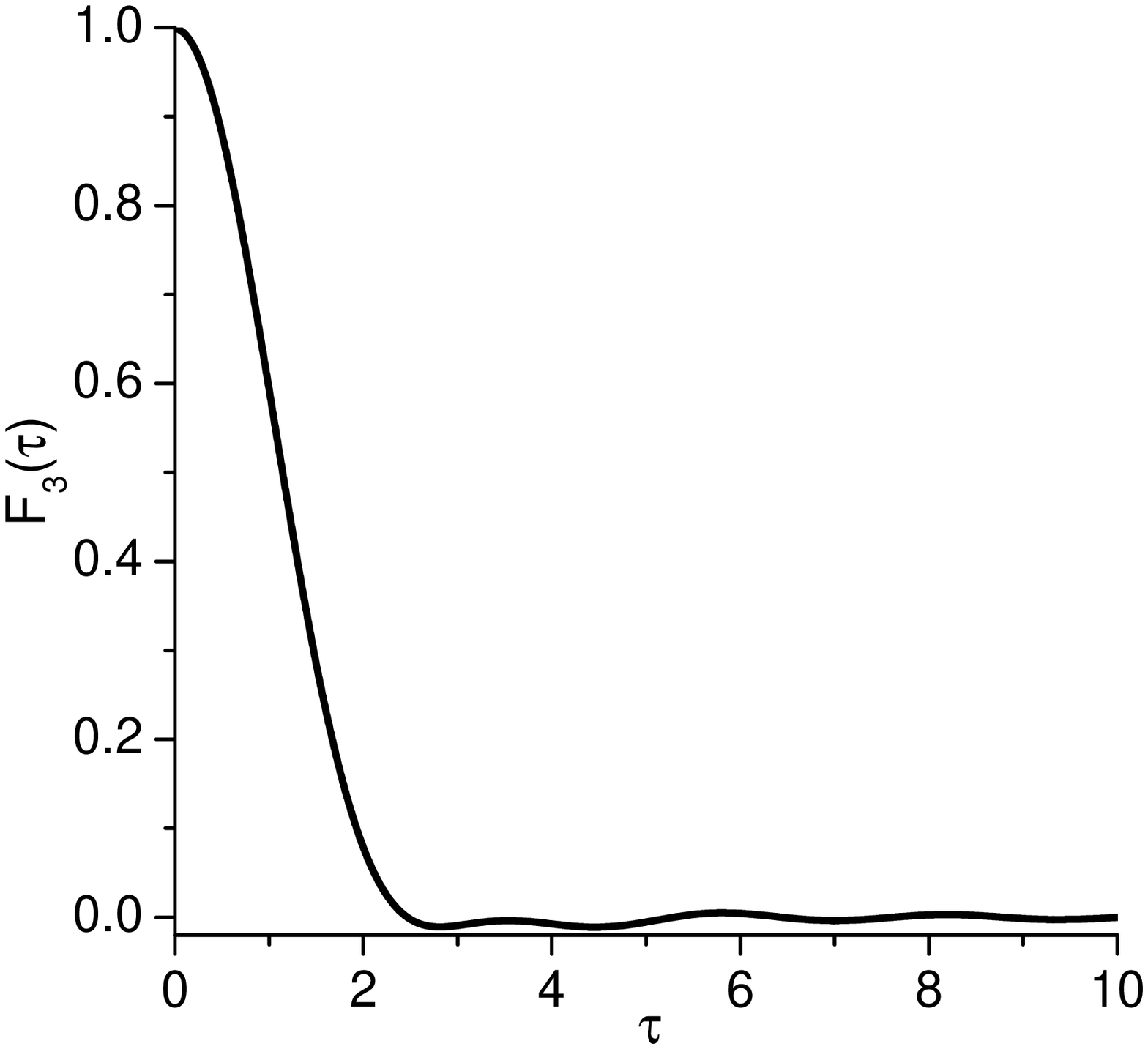}
\includegraphics[height=0.20\textheight,angle=0]{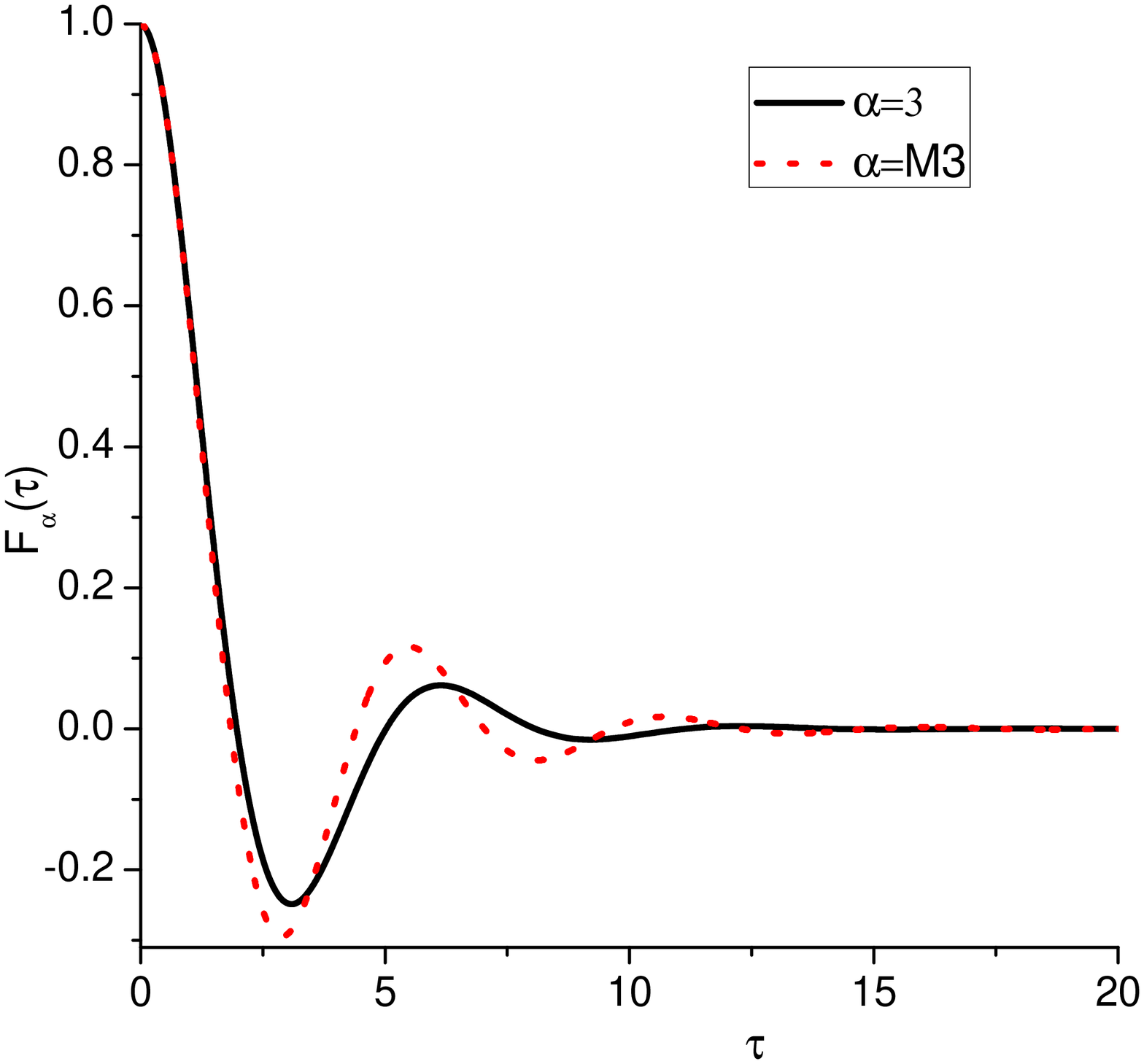}}
\caption{(Colour online) Time evolution of the VAF $F_3(\tau)$
(solid line) at $\Gamma_1\tau^2_{\textrm{LJ}}=1$,
$\Gamma_2\tau^2_{\textrm{LJ}}=1.5$, $\Gamma_3\tau^2_{\textrm{LJ}}=1.05$ (left-hand
panel), $\Gamma_1\tau^2_{\textrm{LJ}}=1$, $\Gamma_2\tau^2_{\textrm{LJ}}=1.5$,
$\Gamma_3\tau^2_{\textrm{LJ}}=1.7$ (central panel) and
$\Gamma_1\tau^2_{\textrm{LJ}}=1$, $\Gamma_2\tau^2_{\textrm{LJ}}=1.5$,
$\Gamma_3\tau^2_{\textrm{LJ}}=4$ (right-hand panel). The dotted line corresponds
to the VAF in the Markovian approximation (labelled by
$\alpha=\text{M3}$).}
\label{fig2}
\end{figure}

The fluid dynamics becomes more diverse if one accepts the
approximation (\ref{G3}) to be valid. In this case, the
corresponding VAFs depend on three SCF $\Gamma_i$, $i=\{1,2,3\}$.
It is seen from the right-hand panel of figure~\ref{fig2} that we can model a
liquid-like dynamics of the fluid with a pronounced minimum of the
corresponding VAF, known as the cage-effect, in addition to the underdamped (solid-like 
dynamics, left-hand panel) and overdamped ones (gas-like dynamics, 
central panel) by varying the value of
$\Gamma_3$ at the fixed parameters $\Gamma_1$ and $\Gamma_2$. This
minimum is related to the back-scattering of the fluid particle
trapped in the cage, which is formed by its neighbourhood. At the
intermediate times, vibration of the particle causes a
rearrangement of the solvation shell, allowing the molecule to
travel away from its initial position, and the oscillations become
completely damped. This case differs from a situation  with the
underdamped oscillations, depicted on the left-hand panel in figure~\ref{fig2},
where one can suggest a more pronounced local ordering in the
fluid, which is capable of trapping the particle several times before it
escapes (thus, the term ``solid-like'' is quite reasonable). At
larger times, the vortex rings in the fluid around a particle can
contribute to the long-time tail modulated by the frequency
$\omega_{\textrm{c}}$ (not shown in figure~\ref{fig2}). Let us also mention that the
Markovian approximation $F_{\text{M3}}(t)$ at large $\Gamma_3$ yields the
results which are very close to those of $F_3(t)$ (see dashed
line in the right-hand panel of figure~\ref{fig2}).

Some words should be said about the nature of the VAFs oscillations
at the long times. First of all, obviously, there is a purely
mathematical reason for them to appear: one should integrate over
the finite interval $[-\omega_{\textrm{c}},\omega_{\textrm{c}}]$ of frequencies when
performing the inverse Fourier transformation from the SFs
(\ref{spectr2}), (\ref{spectr3}) to the VAFs since the
corresponding spectral functions become equal to zero outside this
domain.

On the other hand, let us look at this issue from a physical
viewpoint. The lowest order memory function defines (via the
Green-Kubo relations \cite{MorozovBook}) the generalized
self-diffusion coefficient $\tilde D(\omega)$. In its turn, the
highest order SCF $\Gamma_s$ generates the cut-off frequency, at
which the real part of $\tilde D(\omega)$ vanishes, whereas the
imaginary part of the above mentioned coefficient remains
different from zero. The real part of $\tilde D(\omega)$ is known
to be associated with the energy dissipation by diffusion. Since
$\Re\big[\tilde D(\omega\geqslant\omega_{\textrm{c}})\big]=0$, there appears a
``narrowing'' of the dissipation channel from high
frequencies. At the same time, the imaginary part
$\Im\big[\tilde D(\omega\geqslant\omega_{\textrm{c}})\big]> 0$ contributes to the
vibrational mode of the fluid due to a renormalization of the
corresponding excitation. This underdamped high frequency
excitation can cause a non-monotonous behaviour of the VAFs at long
times.

We are aware that such an explanation cannot be considered as a
complete and irrefragable one. However, to our knowledge, a
commonly accepted viewpoint about a smooth relaxation $\sim
t^{-3/2}$ of the VAF at the hydrodynamic regime is mainly based on
the assumption that correlations in a given dynamic quantity predominantly
decay into pairs of hydrodynamic modes with conserved
variables. This assumption is generally accepted in the MCT. Since
the correlated collisions are considered as a microscopic
precursor of the appearance of long tails, a decay of the excitations
according to the above mentioned scenario yields plane power law
relaxation of the VAFs.

However, in computer simulations it is not possible to
reliably evaluate the VAFs dynamics on timescales exceeding the recurrence
time \cite{Bafile2}, since there appears a spurious overall reduction of the VAF
intensity with some oscillatory and rapidly increasing noisy
behaviour. Though in the recent paper \cite{MHforMemKer}
the authors have managed to obtain the power law behaviour of the
VAF for the supercritical LJ fluid of identical particles from the
direct molecular dynamic simulation, which perfectly agrees with
some theoretical approximations \cite{MHforMemKer1,MHforMemKer2}
for the memory kernels, a question remains still open: 
in what way the VAF tends to its long-time asymptotics. In this context, the
most revealing example is presented in figure~2 of
\cite{MHforMemKer}, where the (negative) hydrodynamic added
mass tends to its asymptotic value non- monotonously.

Besides, we believe that any non-monotonous behaviour of the VAFs
at the long times with vanishing amplitude and the period of
oscillations, which is much smaller as compared to the observation
time, has a little chance to be detected at the experiment in every
detail. Instead, some averaged value (the envelope curve
$at^{-3/2}$) can be only observed.

We also believe (and the preliminary estimations confirm this
assumption) that a more sophisticated ansatz for the memory
kernels, when one postulates the existence of periodical
continued fractions due to relation $\tilde \phi_s(\omega)\approx
\tilde\phi_{s+n}(\omega)$, $n>1$, will allow us to obtain a
smoother behaviour of the VAFs at long times with a less modulated
structure. From a physical point of view, it corresponds to a more
realistic approximation for the VAFs, when relaxation times of the
higher order memory kernels approach their asymptotic values,
alternately, from below and above \cite{our_assump2} rather than
instantly, as it happens within our scheme.

At the end of this section let us discuss another interesting
issue. If we consider the SF (\ref{spectr2}) at
$\Gamma_2=\Gamma_2^{(\textrm{min})}\equiv\Gamma_1/2$, after the inverse
Fourier transformation we obtain another exact result,
\bea\label{Bessel0} F_2(t)\big|_{\Gamma_2=\Gamma_1/2} =J_0(\omega_{\textrm{c}} t), \eea where $J_0$
means the zeroth order Bessel function. It has been shown in a
recent paper \cite{indians} that such a dyna\-mics is typical of
a one-dimensional chain of the particles interacting as harmonic
oscillators. From this point of view, at the threshold value
$\Gamma_2=\Gamma_2^{(\textrm{min})}$, the fluid effectively behaves like a
$1\text{D}$ system with oscillatory power law relaxation $\sim t^{-1/2}$.
Obviously, computer simulations are needed to verify our result by
checking whether a $1\text{D}$ system of
 particles interacting as harmonic oscillators is characterized by
 the equality $\Gamma_2=\Gamma_1/2$.

\section{Transition to the hydrodynamic regime\label{secV}}

In this section, we study the peculiarities of the VAFs transition
to the hydrodynamic regime. It should be emphasized that this
problem is quite topical from both theoretical and applied points
of view. First of all, it is useful to obtain analytical relations
describing in what way (and when) the long-time tails are formed
in the fluid. These expressions (though dependent on the order of
approximation) could be compared with other theoretical methods or
computer simulations data \cite{MHforMemKer} to understand the
microscopic scenario of the hydrodynamic regime formation or even
could be helpful at the interpretation of experimental results.

It is intuitively clear that the time necessary for VAFs to
approach their power law behaviour depends on the level $s$ of
hierarchy, at which the continued fraction [corresponding to the
highest order memory kernel $\tilde\phi_s(z)$] has been summed up:
a greater value of $s$ favors a longer transition time
$\tau_{\textrm H}(\Gamma_1,\ldots,\Gamma_s)$. On the other hand, if one
chooses the value $s$, the problem consists in establishing the
relationship between SCFs $\Gamma_i$ ($i\leqslant s$) and the transition
time.

In the recent papers \cite{russians1,russians2}, the authors have
found the way to evaluate the transition time duration. In the
proposed scheme, a detailed analysis of the SF behaviour on the
complex frequency plane has been performed. The value of the
transition time $\tau_{\textrm H}$ has been shown to be inversely
proportional to the half-width of the gap between branch cuts,
which appear due to the hydrodynamic regime formation of the
corresponding VAFs. The above mentioned discontinuities of the
spectral functions were obtained by Pad\'e-approximation of the VAFs
found from the molecular dynamics and, subsequently, were
continued analytically to the complex frequency domain. Though one
always deals with the finite order  $s$ of the Pad\'e-approximation,
an increase of $s$ means that the number of poles of the spectral
function becomes large enough for discontinuities to mimic some
other types of an irregular behaviour. Strictly speaking, in the
limit $s\to\infty$, an infinite number of poles [see
equation~(\ref{sumZ})] could combine into the discontinuities of some
other structure like branch cuts, responsible for the long time
formation of the corresponding VAFs. Such a picture, though
described rather naively, should stimulate the studies of
collective dynamics of many particle systems based on a more
rigorous mathematical standpoint (see, for example, the discussion
in~\cite{AP}).

To study the transition time dependence on the para\-meters
$\Gamma_i$, like it has been done in
\cite{russians1,russians2}, we analyse the VAFs behaviour on
the complex frequency plane. In contrast to the above cited
papers, our approach allows one to obtain analytical
expressions for $\tau_{\textrm H}(\Gamma_1,\ldots,\Gamma_s)$ due to quite
simple forms~(\ref{spectr2}) and (\ref{spectr3}) of the spectral
functions. Here, we present the results obtained at the assumption
(\ref{G2}), since the case (\ref{G3}) (or even the cases with higher
levels of hierarchy, at which the summed up kinetic kernels are
placed in) can be explored in a similar way.

We start our analysis with a formal classification of the
discontinuities of the Laplace transform
 \bea\label{VAFz} \tilde
F_2(z)=\displaystyle\frac{1}{z+\displaystyle\frac{\Gamma_1}{z+\tilde{\phi}_2(z)}}
\eea of the VAF, where the highest order memory kernel
$\tilde{\phi}_2(z)$  is defined by equation~(\ref{phiS_result}). The
function~(\ref{VAFz}) in the general case can be expanded into the
Laurent series \bea\label{Laurent} \tilde
F_2(z)=\sum\limits_{n=1}^{\infty}\frac
{c_{-n}}{(z-z_0)^n}+\sum\limits_{n=0}^{\infty}c_n(z-z_0)^n.\eea In
equation~(\ref{Laurent}) \bea\label{z0}
z_0=-\frac{\ri\Gamma_1}{\sqrt{\Gamma_1-\Gamma_2}} \eea denotes the
singular point while coefficients of the principal part of the
Laurent series at $\Gamma_1<\Gamma_2$ are expressed as follows:
 \bea\label{cMinus1}
c_{-1}=\frac{\Gamma_1-2\Gamma_2}{2(\Gamma_1-\Gamma_2)}\,, \qquad
c_{-s}\equiv 0\,\, \,\,\,\,\mbox{for all}\,\, \,\,\, s>1.\eea

It follows from the last equality that we deal with a simple pole
of VAF. The pole $z_0<0$ is located on the real axis of the
complex frequency plane and contributes to the VAF in a purely
relaxing manner as $\exp(z_0 t)$. Its effect becomes negligible
at times $t\gg 1/|z_0|$, and the VAF behaviour is completely
defined by the regular part of (\ref{Laurent}), yielding the power
law dependence.

In the case of an underdamped system, when
$\Gamma_2^{(\textrm{min})}<\Gamma_2<\Gamma_1$, the singularity (\ref{z0})
is located at the imaginary axis of the complex frequency plane.
However, contrary to the previous case, $z_0$ turns out to be an
essential singularity point since there are different limits of
(\ref{VAFz}): \bea\label{limits}  \lim\limits_{\epsilon\to
0}\big|\tilde F_2(z_0-\epsilon)\big|=\infty, \qquad
\lim\limits_{\epsilon\to+ 0}\big|\tilde
F_2(z_0+\epsilon)\big|=\frac{\Gamma_2\sqrt{\Gamma_1-\Gamma_2}}{\Gamma_1(2\Gamma_2-\Gamma_1)}\,.
\eea

It can be shown by the numerical estimations that the second limit
in equation~(\ref{limits}) defines the time $\tau_{\textrm H}(\Gamma_1,\Gamma_2)$
needed for the VAF to approach the power law dependence.

Thus, we can summarize the values of transition times to the
hydrodynamic stage of evolution in the following way:
\bea\label{tauS}\left\{\begin{array}{lcc}
\tau_{\textrm H}(\Gamma_1,\Gamma_2)=\displaystyle\frac{A}{|z_0|}+\tau_{\textrm{c}}
&\mbox{at}&
\Gamma_1<\Gamma_2\,,\\[2ex]
\tau_{\textrm H}(\Gamma_1,\Gamma_2)=\displaystyle\frac{B}{|z_0|}\,\,\frac{\Gamma_2}{\Gamma_2-\Gamma_2^{(\textrm{min})}}+\tau_{\textrm{c}}
&\mbox{at}&\, \Gamma_2<\Gamma_1\,,
\end{array}\right.
\eea
 where the parameters $A$ and $B$ depend on the chosen accuracy
of the power law $t^{-3/2}$ approximation of the VAFs. The time
$\tau_{\textrm{c}}=2\piup/\omega_{\textrm{c}}$ is added to match the second maximum of the
kernel (\ref{phiS_t}) at $\Gamma_1\to\Gamma_2$, when the
singularities $z_0$ move to $-\infty$ either on the real axis (at
$\Gamma_1<\Gamma_2$) or on the imaginary axis (at
$\Gamma_2<\Gamma_1$).

It should be emphasized that our estimation for the transition
times (\ref{tauS}) at $\Gamma_1<\Gamma_2$ agrees with the results
of \cite{russians1,russians2} being inversely proportional
to the distance of the singular point from the coordinate origin
in the complex frequency plane. On the other hand, the time it
takes for the VAF to approach the hydrodynamic regime at
$\Gamma_1>\Gamma_2$ cannot be described in the same way, being
renormalized by the factor $\Gamma_2/\big[\Gamma_2-\Gamma^{(\textrm{min})}_2\big]$.
Therefore, the transition time becomes infinitely large when
$\Gamma_2$ tends to its limiting value $\Gamma_2^{(\textrm{min})}$, and at
the point $\Gamma_2=\Gamma_2^{(\textrm{min})}$ there is a crossover from
$t^{-3/2}$ to $t^{-1/2}$-dependence, as it has been already
mentioned in section~\ref{secIV}.

\section{Conclusions\label{secVI}}

In this paper, we study the dynamics of the fluid many-particle
system by an effective summation of the continued fractions,
which correspond to the Laplace transforms of the VAFs.
A cornerstone of our scheme is a physically reasonable hypothesis
\cite{our_assump2,mryglod98} about a convergence of the relaxation
times of higher order memory functions, which are of purely kinetic
origin. The proposed approach has several advantages, namely: i) it
is quite universal being capable of describing the dynamics of a great
number of classical and quantum systems using just one physically
reasonable assumption; ii) it is self-consistent and in a broad
region of parameters reproduces the results of other theoretical
schemes \cite{GCM3,MCT1}; iii) it allows one to model the TCFs'
behaviour with a rather small number of parameters, which can be
taken, for instance, from computer simulations.

An effective summation of the continued fractions for higher order
memory kernels and a subsequent inverse Laplace transformation
bring us to the VAFs that decay at long times as $t^{-3/2}$ and
oscillate. On the other hand, our results in a quite broad domain
of the parameters (which actually are the ``force-force'' static
correlation function and its higher derivatives) reproduce those
obtained by a truncation of the continued fraction at a certain
level $s$ of the hierarchy.

The obtained VAFs have the same zero time moment as well as the
frequency moments up to the order~$2s$ as their counterparts,
obtained within the Markovian approximation for the corresponding
memory kernels. However, in contrast to the results which follow
from the latter case, there are also the finite frequency moments
of all orders while the higher order time moments diverge
(starting from the second moment) due to the
appearance of long tails.

It is also shown that from the viewpoint of Langevin approach we
deal with the Ohmic system. However, there is one essential
distinction: in contrast to a commonly accepted framework
\cite{Leggett}, the obtained spectral weight function sharply
vanishes at the frequency $\omega_{\textrm{c}}$ that yields the observed
power law behaviour of the VAFs.

We show that depending on the hierarchy level, at which the
summed-up continued fraction is placed in, as well as depending on the
relationship between the parameters $\Gamma_i$, one can be faced with
a gas-like, liquid-like, or solid-like behaviour of the
corresponding VAFs. Moreover, at a certain limiting value of the
parameter~$\Gamma_s^{(\textrm{min})}$ there is a crossover from the long
time relaxation $\sim t^{-3/2}$ typical of the hydrodynamic
regime in $3\text{D}$ systems to a slower decay $\sim t^{-1/2}$ of the VAF,
when the fluid behaves like an effective $1\text{D}$~system.

We proposed our explanation of the obtained frequency modulation
of the long tails not only from purely mathematical viewpoint but
also by physical reasoning. A real part of the generalized
self-diffusion coefficient $\tilde D(\omega)$ vanishes at
$\omega\geqslant\omega_{\textrm{c}}$, narrowing the dissipation channel from the
high frequencies; simultaneously, a non-zero value of the
imaginary part of $\tilde D(\omega)$ at $\omega>\omega_{\textrm{c}}$
renormalizes the vibrational (propagating) mode of the fluids. The
above mentioned renormalization with a partially suppressed
dissipation can be a physical reason of slight oscillations of the
VAFs.

In the final part of the paper, we study in detail the
peculiarities of transition to the hydrodynamic regime.
Explicit expressions for times needed for the VAFs to approach the
power law asymptotics are obtained at the assumption (\ref{G2}).
At $\Gamma_1<\Gamma_2$, the transition time is found to be
inversely proportional to the distance of the spectral function
pole from the coordinate origin. At $\Gamma_1>\Gamma_2$, there is
an essential singular  point in the spectral function instead of
a simple pole, and the expression for the transition time is
renormalized, being divergent at $\Gamma_2\to\Gamma_2^{(\textrm{min})}$.
Thus, a general conclusion can be drawn that the hydrodynamic stage
of the fluid dynamics is determined by the regular part of the
Laurent series expansion of the spectral function, whereas the
transition period depends on the nature of singularities.

A natural next step is a description of the real fluids at
particular thermodynamic points rather than mo\-dell\-ing the
VAFs behaviour at some initially prescribed values of the input
parameters. The SCFs $\Gamma_s$ should be calculated by computer
simulations, and the maximal value $s$ of the hierarchy level
should be taken as high as possible. A comparison of the
results calculated within our approach with those obtained by the
MD as well as by the other theories (such as GCM or MCT) could
justify the validity of our scheme. This is a subject of our
forthcoming paper.

\section*{Acknowledgements}

This study was partially supported within the project of the
European Unions Horizon 2020 research and innovation
programme under the Marie Sk\l{}odowska-Curie grant agreement
No 734276.

\ukrainianpart

\title{Простий анзац при дослідженні автокореляційних функцій швидкостей флюїду на різних часових масштабах}
\author{В.В.~Ігнатюк, І.М.~Мриглод, Т.~Брик}
\address{
Інститут фізики конденсованих систем НАН України,
вул. Свєнціцького, 1, 79011 Львів, Україна }

\makeukrtitle

\begin{abstract}
\tolerance=3000%
Запропоновано простий анзац для дослідження автокореляційних функцій швидкостей (АФШ) флюїду на різних часових
проміжках, що базується на ефективному пересумуванні безмежних ланцюгових дробів
на основі фізично обгрунтованого припущення про збіжність часів релаксації ядер вищих порядків, які мають суто кінетичну природу. Отримані в рамках такого анзацу АФШ порівнюються з результатами
марківського наближення для відповідних функцій пам'яті. Показано, що хоча в передемпферованому режимі обидва підходи дають дуже близькі результати на малих та проміжних часах, лише метод пересумовування дробів веде до появи степеневих законів у поведінці АФШ на великих часах, що не спостерігається при описі динаміки флюїду скінченним числом колективних мод. На основі аналізу сингулярностей спектральних функцій на комплексній площині частот отримані явні вирази для оцінки часів переходу від кінетичного етапу еволюції флюїду до гідродинамічного режиму.

\keywords
нерівноважна статистична механіка, статистична гідродинаміка, класичні флюїди, рівняння Ланжевена, марківські процеси

\end{abstract}

\end{document}